**Author for correspondence:**
James A. Klimchuk
e-mail: james.a.klimchuk@nasa.gov

# Key aspects of coronal heating

## James A. Klimchuk

NASA Goddard Space Flight Center, Heliophysics Division, Greenbelt, MD 20771, USA

We highlight 10 key aspects of coronal heating that must be understood before we can consider the problem to be solved. (1) All coronal heating is impulsive. (2) The details of coronal heating matter. (3) The corona is filled with elemental magnetic stands. (4) The corona is densely populated with current sheets. (5) The strands must reconnect to prevent an infinite build-up of stress. (6) Nanoflares repeat with different frequencies. (7) What is the characteristic magnitude of energy release? (8) What causes the collective behaviour responsible for loops? (9) What are the onset conditions for energy release? (10) Chromospheric nanoflares are not a primary source of coronal plasma. Significant progress in solving the coronal heating problem will require coordination of approaches: observational studies, field-aligned hydrodynamic simulations, large-scale and localized three-dimensional magnetohydrodynamic simulations, and possibly also kinetic simulations. There is a unique value to each of these approaches, and the community must strive to coordinate better.

## 1. Introduction

The coronal heating problem—understanding how the upper atmosphere of the Sun is heated to multi-million degree temperatures—remains one of the great unsolved problems in space science. Considerable progress has been made in recent years, as reviewed in [1–3], but a detailed and comprehensive understanding is still lacking. The purpose of this paper is to highlight key aspects of coronal heating that must be explained in order to achieve success. We restrict ourselves to heating of the magnetically closed corona (active regions and quiet Sun). The heating of the open corona (coronal holes and solar wind) is a separate question, possibly with a different answer, if not likely. Chromospheric heating is not specifically addressed, though many of the points made here may be relevant in this cooler part of the atmosphere as well.

There are several different approaches to addressing the coronal heating problem: observational studies; field-aligned hydrodynamic models; large-scale three-dimensional magnetohydrodynamic (3D MHD) models of, e.g., active regions and the global corona; localized 3D MHD or kinetic models of, e.g., individual current sheets and magnetic flux tubes. (Some kinetic physics can be better incorporated into models of all types.) Each approach has its strengths and weaknesses, which must always be kept in mind. Because the approaches are highly complementary, we as community members can benefit greatly by paying closer attention to progress being made in areas different from our own. Encouraging stronger coordination among researchers using different approaches is the main goal of this paper. We highlight 10 key aspects of coronal heating, which take the form of both statements and questions. We hope that they will provide focal points for common discussion and future work.

## 2. All coronal heating is impulsive

All known mechanisms of coronal heating, when applied to a realistic non-uniform corona, predict that the heating is highly time dependent when viewed from the perspective of individual magnetic field lines [1,4]. This includes wave heating as well as reconnection-type heating. The field line perspective is most relevant, because the response of the plasma to the heating is determined primarily by field-aligned processes. The frozen-in flux condition, small plasma $\beta$ (ratio of thermal to magnetic pressures), and efficiency of thermal conduction at transporting energy along but not across the field means that flux strands behave like quasi-rigid, thermally insulated flow pipes. Magnetic field evolution can be important—and is almost certainly important during the coronal heating process—but after the heating has occurred, a one-dimensional hydro description is usually very reasonable.

'Nanoflare' is a term that is often used to describe an impulsive heating event. It was first coined by Parker [5], who envisioned a burst of magnetic reconnection. The meaning of nanoflare has since evolved. Here, as in our earlier work, we take it to mean *an impulsive energy release on a small cross-field spatial scale without regard to physical mechanism*. It is a very generic definition. Waves produce nanoflares (e.g. [1]). Much of the discussion in this paper concerns generic heating, including by waves, but some of it deals specifically with magnetic reconnection. The distinction should be obvious.

Parker's choice of the term nanoflare was motivated by an observation of localized brightenings estimated to contain $10^{24}$ ergs, roughly one billionth that of a large flare. Today's nanoflares can have much lower energy, and they are generally not distinguishable from the many other events occurring along the line of sight.

Finally, we note that nanoflares that recur with a high enough frequency will produce conditions that are similar to steady heating. We return to this important point later.

## 3. The details of coronal heating matter

A common misconception is based on the following chain of reasoning. (i) Photospheric flows slowly stress the coronal magnetic field and inject into it a Poynting flux of energy that is determined by the driver velocity. (ii) The corona adjusts so that the magnetic energy is converted into heat in a statistical steady state (energy out must equal energy in over a long enough time average). (iii) The Poynting flux and therefore the average heating rate do not depend on the details of this adjustment. The first statement is true in part, but it is incomplete. The second statement is certainly true. And the third statement is certainly false, as we now show.

The Poynting flux (erg cm$^{-2}$ s$^{-1}$) associated with the work performed by driver flows in stressing the coronal field is given by

$$F = \frac{1}{4\pi} B_{\mathrm{v}}^2 V_{\mathrm{h}} \tan\theta,$$

where $B_{\mathrm{v}}$ is the vertical component of the field, $V_{\mathrm{h}}$ is the horizontal velocity and $\theta$ is the tilt of the field from vertical in a direction opposite to the motion (the field trails behind the moving footpoint). All quantities are measured at the base of the corona. Consider two cases. In the first, the heating mechanism is highly efficient and operates nearly continuously. Only small stresses can develop in the field. The tilt is small and therefore so too are the Poynting flux and average heating rate. In the second case, the heating mechanism remains dormant for a substantial time and allows stresses to build to sizable levels before switching on. With exactly the same $B_{\mathrm{v}}$ and $V_{\mathrm{h}}$, the Poynting flux and average heating rate are now much larger. The amount of heating depends critically on the mechanism of energy release, i.e. the precise way in which the corona responds to the photospheric driving. Details matter!

Using observed values of $B_{\mathrm{v}}$, $V_{\mathrm{h}}$ and $F$ (equated with the observed energy losses from the corona), we conclude that $\theta$ is approximately $10°$–$20°$ ($10°$ using typical active region values and $20°$ using typical quiet Sun values, though the difference may not be real). $\theta$ is sometimes referred to as the Parker angle, since it was he who first estimated its value [6]. What determines $\theta$? What are the onset conditions associated with the switch-on property described above? This is a fundamental question that has implications far beyond coronal heating.

As a specific example of how the details of heating and the details of a numerical simulation matter, we point to the numerical experiment of Rappazzo *et al.* [7]. They slowly shuffled the footpoints of an initially straight and uniform magnetic field and recorded the total ohmic and viscous dissipation as a function of time. They repeated the experiment several times, changing only the values of the resistivity and viscosity (i.e. the resistive and viscous Reynolds numbers). The footpoint driving was identical in each case. They found that both the magnitude and character of the heating are strong functions of Reynolds number. The time-averaged heating rate increases with Reynolds number, as expected from the above discussion, since stresses can build to higher levels when the dissipation is reduced. Furthermore, the heating becomes much more bursty and nanoflare like. The authors interpret this as turbulence, though a question remains as to whether the turbulence is produced by slow motions in a classical sense, or whether it is the temporary aftermath of instabilities and reconnection events that occur during a quasi-static evolution of the field [8].

## 4. The corona is filled with elemental magnetic strands

The magnetic field of the photosphere is observed to be very clumpy and concentrated into elemental flux tubes. These tubes have a distribution of sizes and distribution of field strengths. The precise shapes of these distributions are not known, but multiple studies reviewed in [9,10] suggest a clustering of values around a characteristic size of about 150 km and characteristic strength of about 1500 G. Most measurements differ from these characteristic values by less than a factor of 2, i.e. are between half and twice as large. The wings of the distributions are uncertain, especially at the low end of the size distribution. At the high end, flux concentrations significantly larger than 150 km are observed, but these appear to be groups of smaller elements. Because most elemental tubes have field strengths near 1500 G, they are often called kilogauss (kG) flux tubes.

The above discussion refers to the net flux that extends upward into the corona. There also exists, at least in the quiet Sun, a weaker field component [9–11]. It has mixed polarity on scales less than 0.5 arcsec and is sometimes called the turbulent field. Only about 10% of the coronal field comes from this component, since most of the turbulent field closes in the photosphere or chromosphere. We concern ourselves here with the kG flux tubes that supply most of the magnetic field in the corona and are implicated in coronal heating, but we note that the weaker turbulent field may play an important role in heating the quiet Sun chromosphere.

Elemental kG flux tubes are confined by the dense photospheric plasma, but they expand rapidly with height to become volume filling in the low-$\beta$ corona. We can estimate the number of tubes, or strands, contained in a single coronal loop by taking the ratio of magnetic fluxes contained in a loop and in a strand:

$$N = \frac{\Phi_l}{\Phi_s} = \frac{B_l}{B_s} \left(\frac{d_l}{d_s}\right)^2,$$

where $B_l$ and $B_s$ are the field strengths and $d_l$ and $d_s$ are the diameters of the loop (measured in the corona) and strand (measured in the photosphere), respectively. For a coronal field strength of 100 G and loop diameter of 1500 km (see §9), we find that a coronal loop contains about seven strands. That is, seven kG flux tubes map into the loop.

Loops are observationally distinct features that tend to draw our attention in a coronal image. However, they account for a relatively small fraction of the coronal plasma [12]. Most plasma is contained in what has been called a diffuse component. Like loops, the diffuse component comprises magnetic strands. A single active region contains upward of 100 000 strands.

The term 'strand' can have different meanings. As used above, a strand is directly linked to the fragmentation of magnetic flux in the photosphere. However, reconnection can cause the field to become much more highly fragmented in the corona. Suppose two strands partially reconnect. We are then left with four topologically distinct features, each with a unique pairing of footpoints. Two will have the same pairings as the parent strands, and two will have entirely new pairings. We can refer to all four of these as individual strands. It is easy to see how multiple coronal reconnections can lead to many strands emanating from a single flux concentration in the photosphere. Yet another usage of the term strand is based on the plasma properties. A plasma strand, defined to have approximately uniform temperature and density over a cross section, could be a subset of a fatter magnetic strand. This is not likely to be a common occurrence, however, at least for reconnection-based heating processes. Temperature and density are determined by the heating, and a magnetic strand that is created by reconnection will tend to have uniform heating over its entire cross section. The meaning of the term strand must be taken from the context in which it is used.

## 5. The corona is densely populated with current sheets

Photospheric flows associated with the evolving pattern of turbulent convection act upon the footpoints of strands. Translational motions cause the strands to become tangled about each other, and rotational motions cause the strands to become individually twisted (figure 1). Parker [13] suggested that infinitely sharp current sheets (tangential discontinuities) must be produced in the corona, even when the magnetic flux and the flow are both continuous in the photosphere. This suggestion has been actively debated [14]. But even if true discontinuities do not occur as envisioned by Parker, narrow sheets of intense current are a necessary consequence of the fragmentation of the field in the photosphere (both the fragmentation of the flux and the fragmentation of the field line connectivity) (e.g. [15]). As depicted in figure 1, abrupt rotations of the field are expected across the boundaries between strands, even if the axes are perfectly aligned. This implies current sheets.

An active region that contains 100 000 or more strands will contain a similar number of current sheets. No present-day MHD simulations are capable of resolving this ubiquitous fine-scale complexity. Is this important? Does coronal heating involve these current sheets in a fundamental way? If so, does the mechanism that provides the heating in a numerical simulation (e.g. Ohmic dissipation on a much larger scale) mimic the properties of the real mechanism that operates on the Sun? Does it have the correct switch-on properties? These are questions that must ultimately be answered.

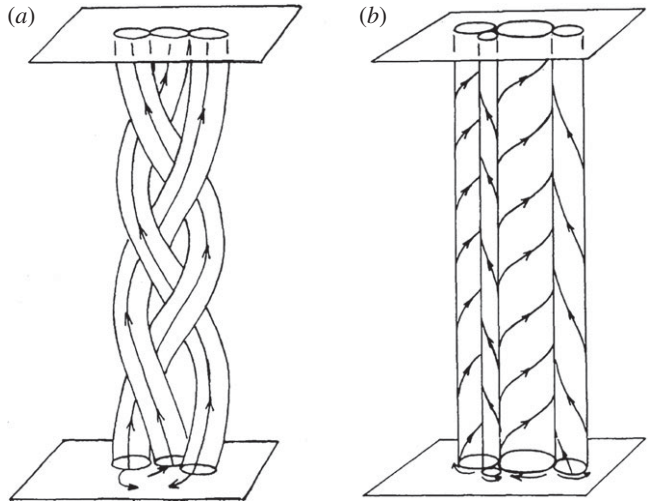

**Figure 1.** (*a*) Tangled and (*b*) twisted magnetic strands. Reproduced with permission from [6].

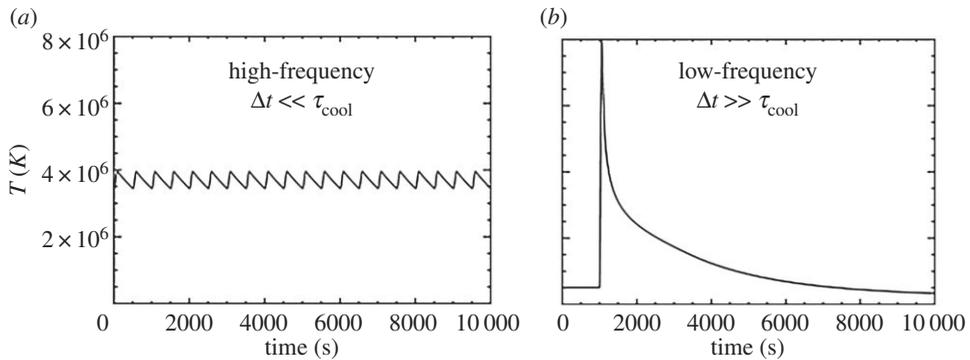

**Figure 2.** Temperature evolution of a coronal strand heated by (*a*) high-frequency and (*b*) low-frequency nanoflares.

## 6. The strands must reconnect to prevent an infinite build-up of stress

Photospheric observations show that tangling and twisting of the coronal field must be occurring. The field must therefore reconnect in order to prevent an infinite build-up of stress. This produces unavoidable plasma heating, even if other forms of heating, e.g. wave based, are also occurring. What are the relative magnitudes of the different types of heating?

## 7. Nanoflares repeat with different frequencies

As already mentioned, an important aspect of nanoflares is the frequency with which they repeat on a given strand. This has a strong influence on both the instantaneous and time-averaged properties of the plasma. What matters specifically is the heating frequency with respect to the plasma cooling time. Figure 2 shows the temperature evolution of a strand that is heated at two different frequencies. This is the average temperature along the strand as computed with the Enthalpy Based Thermal Evolution of Loops (EBTEL) code [16,17]. On the left, the delay between successive nanoflares is much less than a cooling time, so the temperature fluctuates about a mean value. This is referred to as high-frequency heating. On the right, the delay is much longer than a cooling time, and the strand cools fully before being reheated. This is low-frequency heating.

Many if not most observationally distinct coronal loops are well explained by a 'storm' of low-frequency nanoflares. Each strand of the loop bundle is heated once during the loop lifetime, but different strands are heated at different times [18,19]. Shorter duration storms produce loops that are more nearly isothermal at any given instant than do longer duration storms, i.e. the loops have a narrower temperature distribution. Nanoflare storms can explain the observed temperature-dependent pattern of under and over density relative to static equilibrium, as well as the observed delay in peak emission seen with different channels of the Atmospheric Imaging Assembly (AIA) on the Solar Dynamics Observatory (SDO) [12]. The delay between peak emission observed in soft X-rays by Yohkoh and EUV emission by TRACE may be more difficult to explain [20].

An alternative explanation of observationally distinct loops is thermal non-equilibrium, which occurs when steady heating is highly concentrated near the loop footpoints [21]. This interpretation has not yet been as rigorously tested as nanoflare storms. Furthermore, thermal non-equilibrium conditions cannot persist for the entire loop evolution, since that would produce loop properties that are inconsistent with observations [22]. We suggest that upward and downward displacement of the chromosphere in response to changing coronal pressure, combined with a strong height dependence of the heating, leads to an oscillation between thermal non-equilibrium conditions and steady flow conditions. This allows for loop cooling but prevents the formation of cold condensations and other properties that are at odds with loop observations. Coronal rain is an exception and can be explained by thermal non-equilibrium alone.

It is much more challenging to determine the nanoflare frequency in the diffuse component of the corona than it is in distinguishable loops. One method is to study the slope of the emission measure (EM) distribution coolward of its peak. Steep slopes imply high-frequency heating, and shallow slopes imply low-frequency heating. A number of investigations have measured the slope in the cores of active regions, which are generally diffuse. A review of the published values and model predictions can be found in [23]. Different active regions seem to have different slopes and therefore different nanoflare frequencies. It must be remembered, however, that the observed slopes have large uncertainties [24]. Cargill and co-workers [25,26] recently showed that the full range of observed slopes can be understood if (i) nanoflares occur with a random distribution of energies that obey a power law, (ii) there is a correspondence between the event size and the delay between successive events, and (iii) the mean delay is of order 1000s, comparable to a cooling time.

Viall & Klimchuk [27] have used a different approach to study nanoflare frequencies in the diffuse corona. They developed an automated procedure to measure time lags in the brightness variability observed in different channels of SDO/AIA. A large majority of pixels in the images show clear evidence of cooling plasma, presumably in the aftermath of nanoflares, though not all the plasma along the line of sight need be cooling. The time lags are seen even in channels with widely separated temperature, implying that the nanoflares have low frequency. The fact that a clear signal can be detected seems rather amazing when one realizes that the emission in each pixel comes from many hundreds to thousands of unresolved strands. Modelling shows that the technique is expected to see a cooling signature even with such extreme averaging [28].

Intermediate to high-frequency heating is needed to explain active region cores with steep EM slopes. Low-frequency heating is needed to explain active region cores with shallow EM slopes as well as the ubiquitous time lags. Both types of heating can be present along any given line of sight, and the relative mix of frequencies can vary from one line of sight to the next and from one active region to the next. The challenge is to explain both the slope and the time lag within the same framework, possibly as proposed by Cargill and co-workers [25,26].

## 8. What is the characteristic magnitude of energy release?

Physical understanding of what determines nanoflare frequencies is a crucial goal. An obvious factor in determining the frequency is the magnitude of energy release. If each event extracts a large portion of the available free magnetic energy in the field, more time will be needed for photospheric flows to rebuild the stress. Using the schematic drawing of two misaligned strands in figure 3, we can estimate the energy release in a single reconnection event. The strands are pressed against each other and have a finite surface of contact. When field lines reconnect near the centre of this surface, they pull away from the reconnection site as indicated. There is a limit, however, as to how far they can retract. Only two flux tubes are shown in the figure, but the low-β corona is completely filled with magnetic field. When the retracting field lines reach the edge of the contact area, they are prevented from moving much farther by other field lines not shown.

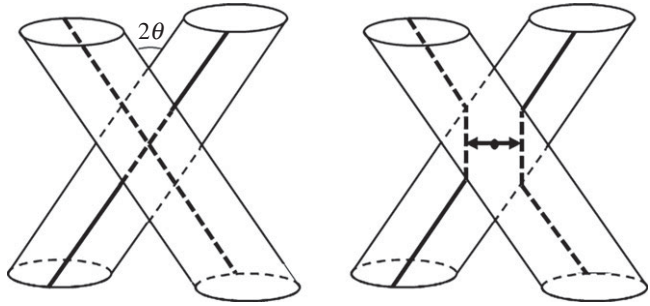

**Figure 3.** Reconnecting magnetic field lines at the interface between two misaligned strands.

The energy released by the reconnection is roughly proportional to the decrease in the length of the field lines, since the strength of the field is largely unchanged. From simple geometric considerations, we find that the energy release per unit cross-sectional area of the reconnecting flux is

$$\frac{\Delta E}{A} = \frac{1}{8\pi} r B^2 \left( \frac{\sec \theta - 1}{\sin \theta} \right),$$

where $r$ is the radius of the strands, $B$ is the field strength and $\theta$ is the half angle between the strands (corresponding to the tilt discussed earlier). A typical 1500 G flux concentration in the photosphere has a radius of 75 km. From conservation of flux, the corresponding radius in the 100 G corona of an active region is 290 km. For $\theta = 10°$ from the Poynting flux arguments in §3, the amount of released energy is then $1.0 \times 10^9$ erg cm$^{-2}$. Using 10 G, 920 km and 20° for the quiet Sun gives $6.9 \times 10^7$ erg cm$^{-2}$. Comparing these values with the canonical energy loss rates of $10^7$ and $3 \times 10^5$ erg cm$^{-2}$ s$^{-1}$ for active regions and quiet Sun [29], we conclude that the average delay between reconnection events in a given strand is roughly 100 s and 230 s, respectively. It should be stressed that these are only very approximate estimates. We can expect a broad distribution of reconnection energies and delay times given the wide range of field strengths in the corona. Furthermore, the derived values should be treated as upper limits for a given $B$ since the reconnecting strands may be more fragmented (have smaller radii) than implied by kG flux concentrations in the photosphere, as discussed in §4.

Twisted strands will also reconnect, even if they are untangled. When they do, there is a partial unwinding of the field, and the decrease in energy is given approximately by

$$\frac{\Delta E}{A} = \frac{1}{24} r^2 B^2 \frac{\varphi}{L},$$

where $\varphi$ and $L$ are the twist and axial length. Twist is limited by the kink instability to a maximum value of roughly $5\pi$ [30–32]. Using $L = 5 \times 10^9$ cm and the above values for $r$ and $B$, we get an upper limit for the delay between events of roughly 110 s in active regions and 370 s in the quiet Sun. Note that the pitch angle of the field (tilt from vertical) is related to twist according to $\theta = \arctan(\varphi r/L)$. With the above values, $\theta = 5°$ and $16°$ in active regions and quiet Sun, respectively, which is comparable to the Parker angle (§3).

An average delay of 100 s is in the high-frequency regime. However, since the response of the plasma depends only weakly on the details of the heating temporal profile, multiple reconnection events can cluster together to produce what is effectively a single nanoflare. This is indicated schematically in figure 4. More realistic examples can be found in [26,33]. Nanoflares can therefore occur with low frequency even if the average delay between reconnection events corresponds to a high frequency. Understanding the clustering of events is a critical aspect of the coronal heating problem. Events could cluster by chance or because of a physical connection such as an avalanche-like process where one event triggers subsequent events.

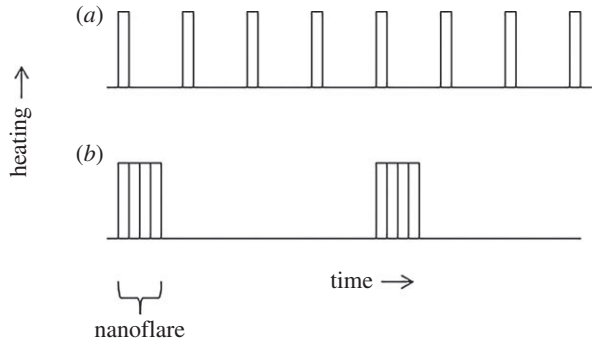

**Figure 4.** Heating rate versus time for low-energy, high-frequency nanoflares (*a*) and high-energy, low-frequency nanoflares that result from clustering (*b*).

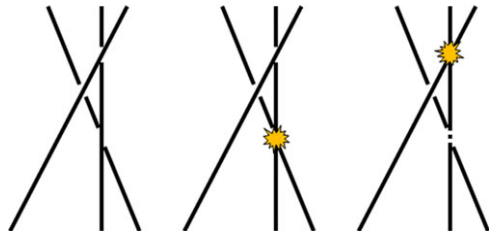

**Figure 5.** One magnetic strand reconnecting with two others. (Online version in colour.)

It is important to realize that reconnection events which combine to form a single nanoflare need not occur at the same location. Figure 5 shows a schematic example. The vertical strand first reconnects with the strand behind and then with the strand in front. The evolution of the plasma is largely independent of where along the strand the heating occurs [34]. Only the very earliest phase is affected by the location. We have little observational information about this phase, because the plasma is exceptionally hot and faint and therefore difficult to detect. This is unfortunate, because the initial response of the plasma holds tremendous clues about the nature of the energy release [35]. The situation could change for the better with future missions like Solar-C and rocket experiments like the Extreme Ultraviolet Normal Incidence Spectrometer (EUNIS), Marshall Grazing Incidence X-ray Spectrometer (MaGIXS) and VEry high angular Resolution Imaging Spectrometer (VERIS). EUNIS has already detected evidence for pervasive Fe XIX emission from an active region [36].

When strands reconnect, they exchange segments. If the strands have similar plasma properties before reconnection, then they will evolve in a similar manner after reconnection. However, if the initial properties are different, then the evolution will also be different. This is indicated schematically in figure 6. Red and blue represent different temperatures and/or densities. One of the new strands formed by reconnection comprises a short red segment and a long blue segment. The plasma mixes and becomes deep purple. The other strand comprises a long red segment and a short blue segment, which mix to become lavender. Field-aligned hydo models do not generally take this mixing into account, though they could, as in the case of Bradshaw *et al*. [37]. Separate simulations can be spliced together at the time of the impulsive energy release attributed to reconnection. The effects are only important for high-frequency heating. With low-frequency nanoflares, the evaporated plasma dominates over the pre-existing plasma in the strand.

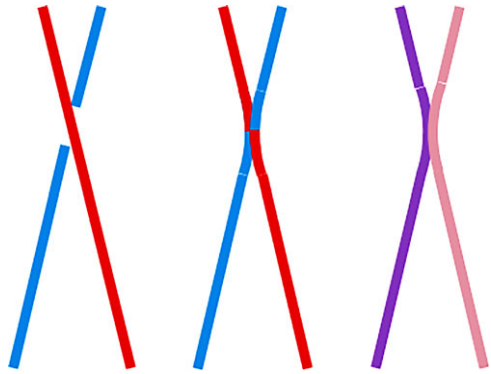

**Figure 6.** Two reconnecting strands filled with different plasmas.

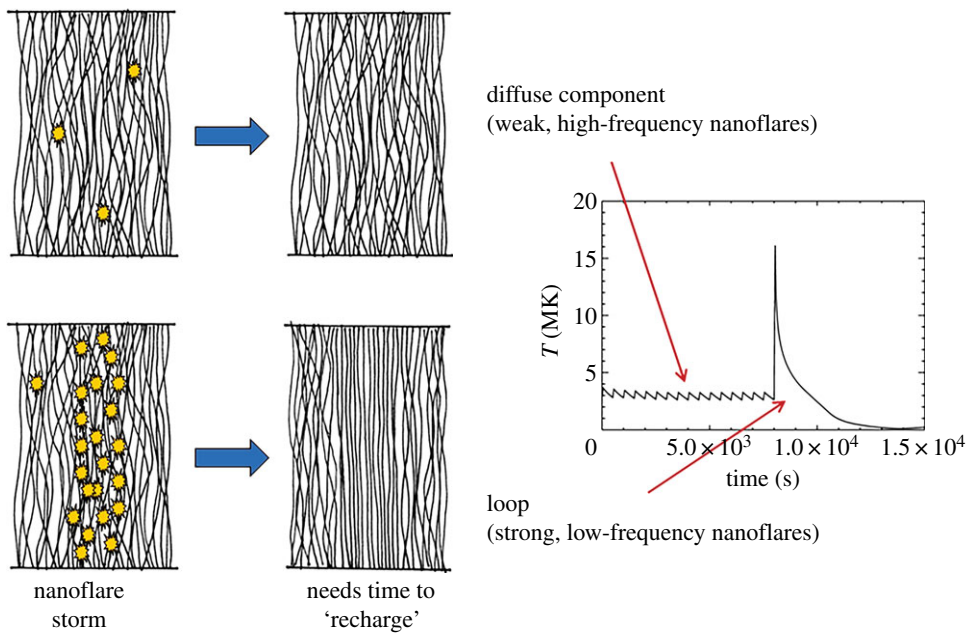

diffuse component
(weak, high-frequency nanoflares)

loop
(strong, low-frequency nanoflares)

nanoflare
storm

needs time to
'recharge'

**Figure 7.** A unifying picture that explains both the diffuse component of the corona and distinct loops.

Figure 7 suggests a unifying picture that attempts to explain both the diffuse component of the corona and observationally distinct coronal loops. Over much of the corona, nanoflares occur at intermediate to high frequency, and temperature variations are small. Footpoint driving replenishes the magnetic energy extracted by each modest event in near real time, producing a statistical steady state. This is the diffuse corona. (The diffuse cores of some active regions may be better explained by low-frequency nanoflares, as discussed in §7.) At certain places, the occurrence of reconnection events increases dramatically. They cluster to form highly energetic nanoflares, and temperatures increase tremendously. Because much more energy has been extracted from the field, it takes more time for footpoint driving to rebuild the stress, i.e. the 'recharging' time is longer. The plasma is able to cool fully before subsequent nanoflares return. This is the low-frequency heating that produces distinct loops. Note that some authors use the term 'unresolved corona' instead of diffuse component. We avoid this because the diffuse component and distinct loops are both composed of spatially unresolved strands.

This scenario suggests the existence of critical conditions for energy release, such as a critical misalignment angle between adjacent strands. Lopez Fuentes & Klimchuk [33,38] have studied the consequences of a critical angle with a cellular automaton model. They find only minimal clustering of reconnection events (rarely more than three events in a single nanoflare), but this could be due to the simplicity of the model. Further work is underway.

## 9. What causes the collective behaviour responsible for loops?

Whether or not the above picture is correct, the existence of distinguishable loops suggests a collective behaviour that is not present in the diffuse corona. What is the origin of this collective behaviour? Are all events activated by a single source, or is there an avalanche-like process where one event triggers another, which triggers another, etc.? What determines the cross-field spatial scale of the nanoflare storm, i.e. what determines the diameters of loops?

Loops seem to have a characteristic diameter of approximately 1500 km. More precisely, 1500 km is one preferred spatial scale that appears in the measurements. There may be other preferred scales, such as the diameters of the unresolved strands that comprise loops, but the envelope of the strand bundle is often around 1500 km.

A common method to measure the diameter of a loop is to determine the width of the cross-axis intensity profile after subtracting a background. The preferred way to do this, which does not depend on any assumptions about the shape of the profile, is to compute its standard deviation [39]. If the loop cross section is circular and radiates uniformly, then the diameter is four times the standard deviation.

The typical diameter of 1500 km measured from TRACE and SDO/AIA images is not much bigger than the instrument resolution. Although the effects of telescope point spread function (PSF), detector pixelation, and background were taken into account in some of the studies [40–42], the veracity of the measurements remained in question [43]. The recent flight of the Hi-C rocket experiment provided an outstanding opportunity to address this issue, since it has a spatial resolution three to six times better than AIA [44]. We report here on the initial results of a comparative study of Hi-C and AIA measurements (team led by C. DeForest). The author (J.A.K.) identified four loops in an AIA 193 A image obtained at the same time as the rocket flight. The loops were selected based on the same subjective criteria used in numerous previous studies: a substantial portion of the loop must be free of a complicated background in order to allow for the study of loop expansion. Widths were measured using the method described above, first with the AIA image, then with the corresponding Hi-C image, and finally with a Hi-C image degraded to match the resolution of AIA.

Figure 8 shows the Hi-C 193 A image with the four loops marked. Loops 2, 3 and 4 can be seen in greater detail in the figs in [45]. Figure 9 shows the measured standard deviations (widths) as a function of position along the loop, in units of Hi-C pixels (0.103 arcsec or 75 km). The solid, dashed and dotted curves are the Hi-C, AIA and degraded Hi-C measurements, respectively. The three measurements are similar overall. Hi-C values are systematically smaller, but by generally less than 25%. This is likely to be due to the coarser resolution of AIA, which may be different from previously thought, as evidenced by the differences in the AIA and degraded Hi-C measurements. The important point is that the Hi-C widths are not several times smaller than the AIA widths, which would be the case if the loops were highly under resolved by AIA. We conclude that the widths measured in previous studies using TRACE and AIA data are essentially accurate. The four loops presented here are part of a larger study that will be reported in a separate paper.

There have been multiple published studies of loop widths, but they use different measures of the width, so to make a comparison it is necessary to convert to a common measure. As stated above, a circular cross section of uniformly emitting plasma produces a cross-axis intensity profile with a standard deviation that is one-fourth the loop diameter. The standard deviation of a Gaussian profile, which is often assumed for making fits, is equal to the Gaussian half width, which is 0.42 times the full width at half maximum (FWHM). To compare measurements, we convert to loop diameter, which we define to be four times the standard deviation and 1.70 times the FWHM.

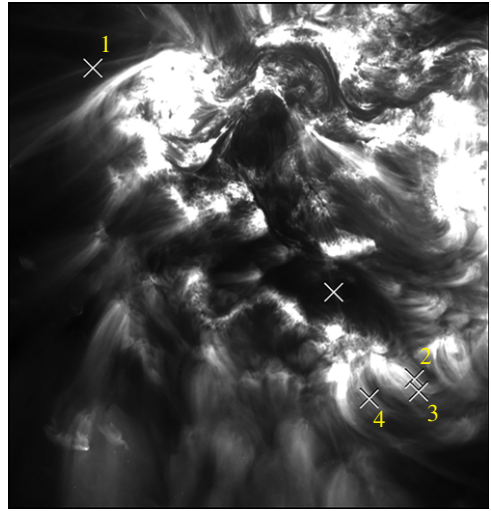

**Figure 8.** Hi-C image in 193 A showing the four loops selected for width measurements.

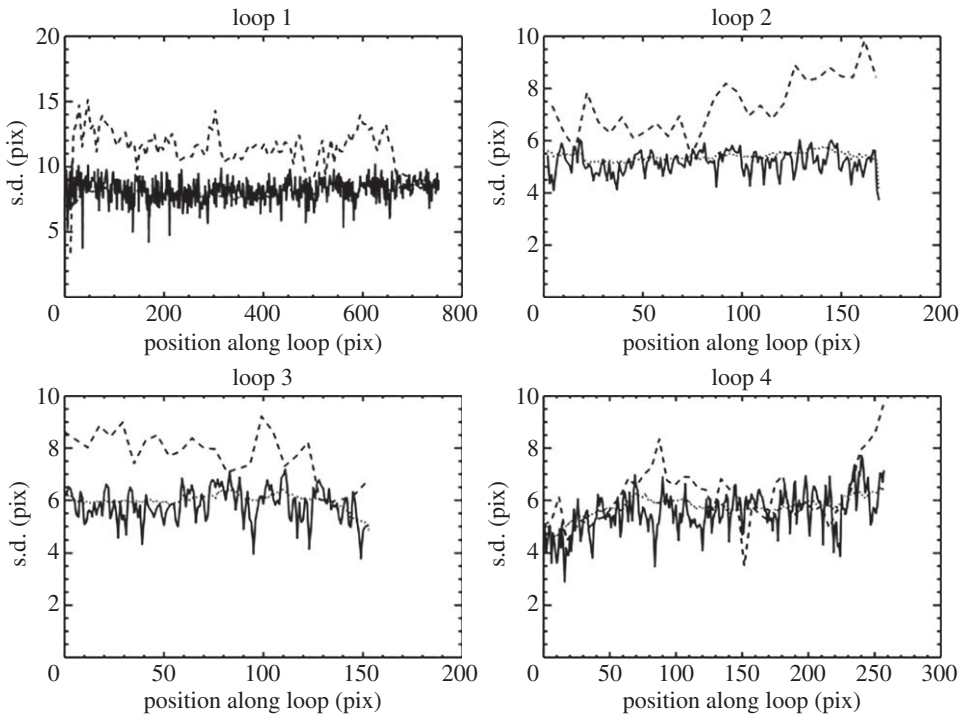

**Figure 9.** Measured standard deviation (width) versus position along the four loops marked in figure 8. Hi-C (solid), AIA (dashed) and degraded Hi-C (dotted) images were used for the three sets of measurements. Units are Hi-C pixels. The diameter is approximately four times the standard deviation.

The 4 Hi-C loops in figure 9 have diameters between 1500 and 2400 km. The 20 TRACE loops in [41] have diameters mostly in the range 900–1800 km when corrected for the PSF (approx. 30% larger when uncorrected). The 78 TRACE loops in [46] have uncorrected diameters mostly in the range 1800–3000 km with a mean of 2400 km. The 91 Hi-C loop segments in [47] have diameters mostly in the range 400–2000 km. Morton & McLaughlin [48] report one Hi-C loop with a diameter of 920 km and a second bundle of threads having individual diameters of 600–1240 km.

Taken together, these studies suggest a preferred spatial scale near 1500 km. Many observed loop diameters cluster near this value to within a factor of about 2. Note that these results apply explicitly to 'warm' loops with temperatures near 1–2 MK. Hot loops do not necessarily obey this trend, though see [49].

Morton & McLaughlin [50] have measured other elongated structures with Hi-C that are not traditional coronal loops. They appear within 'dark inclusions' associated with bright, reticulated moss. The structures have lengths of a few thousand kilometres and are interpreted as the lower transition region ends of hot loops that extend downward below the bright moss into the chromosphere. The diameters are mostly in range 340–1200 km with a mean of 750 km. The relationship of these structures to coronal loops requires further investigation.

As already emphasized, the existence of a preferred scale near 1500 km does not preclude the possibility that loops have other preferred scales of smaller size that are below our resolving capability. Brooks *et al.* [51] used densities and loop widths measured by the Extreme ultraviolet Imaging Spectrometer (EIS) on Hinode to infer the existence of substructure. They estimated the minimum number of strands that must be present within each of 20 loops and found that only a few (three to eight) strands are generally required. It must be remembered that this is a lower limit and that many more strands are also consistent with the data. Winebarger *et al.* [52] studied intensity variations in nearby Hi-C pixels and concluded that they are consistent with photon counting statistics, suggesting that there is little structure on the scale of a Hi-C pixel (approx. 75 km). Moss regions and regions of sheared magnetic field are an exception, and structure on a sub-pixel scale is not precluded. Finally, [45] concluded that the smoothness of cross-axis intensity profiles in Hi-C loops implies that each loop must contain at least 7500 strands of diameter no greater than 15 km. This lends support to our discussion in §4 that partial reconnections between flux tubes can greatly multiply the number of topologically distinct coronal strands in comparison to the number of elemental photospheric flux tubes.

## 10. What are the onset conditions for energy release?

We have emphasized that the mechanism of magnetic energy release must not switch on until significant magnetic stresses have developed. Otherwise, the magnitude of energy release would be small. This is true not only for coronal heating, but also for flares, coronal mass ejections, jets, spicules, etc. What are the onset conditions responsible for this crucial switch-on property?

One possibility is that significant energy release does not occur until the magnetic fields are sufficiently misaligned. The secondary instability of current sheets is an explosive instability that occurs whenever the rotation of the field across the sheet exceeds a critical value of approximately 40° [53,54]. If we equate this misalignment angle with twice the tilt angle of the field on either side of the sheet, then there is a reasonable agreement with the observationally inferred Parker angle. We imagine that footpoint driving steadily increases the tangling and twisting of magnetic strands until the critical angle is reached, at which point the instability sets in.

Onset of reconnection might also depend on the thickness of the current sheet. One scenario is that current sheets reconnect at a very slow Sweet–Parker rate until they become sufficiently thin that 'Hall terms' in the generalized Ohm's Law become important. At that point, there is a sudden transition to a fast Petschek-like reconnection [55,56]. The sheet must be extremely thin, however (ion skin depth without a guide field or ion gyro-radius with one). New simulations show that fast reconnection can in fact occur at a much greater thickness, but only if the aspect ratio of the current sheet (ratio of length to width) is at least 100 [57,58]. This suggests a scenario in which footpoint driving decreases the thickness of current sheets, by enhancing the surrounding magnetic pressure, until a critical thickness or aspect ratio is reached.

Yet another possibility is a critical twist associated with the kink instability [30–32]. This could occur at the level of a strand or of an entire loop. When a twisted loop undergoes resistive internal kinking, the smoothly distributed currents that are present before kinking are rapidly converted into a large number of current sheets that are scattered throughout the loop volume [59]. These sheets can then reconnect to heat the plasma. This is an example where the individual events in a nanoflare storm are activated by a single source, rather than being the result of an avalanche (although the formation of multiple current sheets is not yet fully understood and may have an avalanche-like quality). Another interesting aspect is that the kinking of one loop, or strand, can trigger the kinking of another nearby [60].

## 11. Chromospheric nanoflares are not a primary source of coronal plasma

Heating is expected to occur at all heights in the solar atmosphere, including but not limited to the corona. In fact, the heating requirements of the chromosphere are much larger than those of the corona [29]. This suggests the interesting possibility that coronal plasma comes not from heating in the corona but from heating in the chromosphere. If chromospheric plasma is directly heated to coronal temperatures and ejected upward, might this explain much of what we observe in the corona?

We have looked at this question carefully and concluded that the answer is no [61–64]. When chromospheric plasma is heated impulsively to, e.g. 2 MK, its pressure suddenly increases by two orders of magnitude. This leads to an explosive expansion upward along the magnetic field at several hundred kilometres per second. The expanding plasma cools at a tremendous rate, simply from the work done by expansion, i.e. even if the expansion is adiabatic and excludes radiative and conductive cooling. We have computed spectral line profiles of Fe XII (195) and Fe XIV (274) expected from this scenario and find that, in comparison to actual observations, (i) the intensities are much too faint, (ii) the blue shifts are much too fast, (iii) the blue-red line asymmetries are much too large, and (iv) the emission is confined to low altitudes (less than 10 Mm). We conclude that chromospheric nanoflares, including those that might heat the tips of type II spicules, are not a primary source of coronal plasma. (See [65] for an alternative explanation for the hot plasma at the tips of spicules.) Chromospheric nanoflares may play a major role in powering the radiation from the chromosphere, and they may generate waves that are important for heating the corona, but they do not, in general, raise the temperature of chromospheric plasma to coronal values.

A concern has been raised about the generality of the above studies because they impose a spatially localized nanoflare in the chromosphere. Actual heating may be more broadly distributed with height. While this is true, it should not negate the conclusions. The energy per unit volume of an impulsive heating event can be roughly equated with an increase in pressure. As long as the heating is significantly greater in the chromosphere than in the corona, there will be a large pressure differential and an explosive expansion and cooling of the plasma, which we have shown is inconsistent with observations. Furthermore, we can rule out a comparable level of impulsive heating in the corona and chromosphere, since it would raise the temperature of the much lower density corona to more than 100 MK (if the chromosphere is heated to 2 MK). Such extreme temperatures are not observed. Even if this were to occur, most of the coronal plasma observed at traditional temperatures would come from evaporation, not from direct chromospheric heating, which was our point in the first place. We affirm the conclusion that most coronal plasma is the result of heating that occurs in the corona. But we also recognize that this view is not universally accepted.

## 12. Conclusion

We have highlighted 10 key aspects of coronal heating that must be addressed and understood before we can consider this long-standing problem to be solved. Progress would be greatly enhanced with an improved coordination of approaches: observations, field-aligned hydro simulations, large-scale and localized 3D MHD simulations, and kinetic simulations [66]. We hope that, by highlighting these key aspects, we will help to facilitate this coordination.

Acknowledgements. I thank the organizers of the New Approaches in Coronal Heating Discussion Meeting, Ineke De Moortel and Philippa Browning, for the opportunity to present and discuss this work. Numerous people have contributed in one way or another to the ideas presented here. Special thanks to Peter Cargill. I also thank the referees for helpful suggestions that improved the presentation.

**Funding statement.** This work was supported by the NASA Heliophysics Guest Investigator and Supporting Research and Technology Programs.